\documentclass[12pt,oneside,a4paper]{article}

\usepackage[english]{babel}
\usepackage{amsmath,amssymb,amsfonts}
\usepackage[gen,right]{eurosym}
\usepackage{textcomp,longtable,epsfig,url}
\usepackage{bbm}
\usepackage{graphicx}
\usepackage{subfigure}
\usepackage{hyperref}
\usepackage[intoc]{nomencl}
\usepackage[printonlyused]{acronym}
\usepackage{verbatim}
\usepackage{tabularx}
\usepackage{url}
\usepackage{color}
\usepackage{amssymb,amsmath}
\usepackage{setspace}
\usepackage{scrhack}
\usepackage{floatrow}
\usepackage{listings}
\usepackage{amsthm}
\usepackage{paralist}
\usepackage{xr}
\usepackage{graphicx}
\usepackage{listings}
\usepackage{booktabs}
\usepackage{flafter}
\usepackage{varwidth}
\usepackage[round]{natbib}
\usepackage[a4paper]{geometry}
\usepackage{enumitem}
\usepackage{titling}

\makeindex

\newtheorem{definition}{Definition}[section]

\newtheorem{theorem}[definition]{Theorem}
\newtheorem{example}[definition]{Example}

\newcommand{\beqo}{\begin{eqnarray*}}
	\newcommand{\eeqo}{\end{eqnarray*}\noindent}
\newcommand{\beq}{\begin{eqnarray}}
	\newcommand{\eeq}{\end{eqnarray}\noindent}

\title{Some remarks on the effect of risk sharing and diversification for infinite mean risks}

\author{ Alfred M\"uller\\ Department Mathematik\\ University of Siegen, Germany\\ \texttt{mueller@mathematik.uni-siegen.de}}

\date{\today.\\[.3cm]
	}

\begin{document}
\begin{titlingpage}
		\maketitle
\begin{abstract}
The basic principle of any version of insurance is the paradigm that exchanging risk by sharing it in a pool is beneficial for the participants. In case of independent risks with a finite mean this is the case for risk averse decision makers. The situation may be very different in case of infinite mean models. In that case it is known that risk sharing may have a negative effect, which is sometimes called the nondiversification trap. This phenomenon is well known for infinite mean stable distributions.  In a series of recent papers similar results for infinite mean Pareto and Fr\'echet distributions have been obtained. We further investigate this property by showing that many of these results can be obtained as special cases of a simple result demonstrating that this holds for any distribution that is more skewed than a Cauchy distribution. We also relate this to the situation of deadly catastrophic risks, where we assume a positive probability for an infinite value. That case gives a very simple intuition why this phenomenon can occur for such catastrophic risks. We also mention several open problems and conjectures in this context. 

~\\			
			
{\raggedleft {\bf Keywords:} risk sharing, infinite mean models, nondiversification trap, skewness}   \\[0.2cm]
					\end{abstract}
	\end{titlingpage}

	\pagestyle{plain}
	
	\onehalfspacing
	\allowdisplaybreaks

\section{Introduction}

The basic reason for the existence of any kind of insurance or other versions of risk sharing is the fact that risk sharing typically leads to a reduction of risk. Under the assumption of independent and identically distributed risks with a finite mean it is a simple consequence of the law of large numbers that sharing risks will typically be beneficial for risk averse decision makers. The problem of an adverse effect of diversification in case of infinite mean distributions has already been described in a seminal papers by \cite{fama1965} and \cite{samuelson1967} for the case of stable distributions. 
Further studies of this problem with infinite mean stable distributions have followed, e.g. by \cite{ibragimov2009}, where the term \textit{nondiversification trap} has been introduced for the situation, where an individual does not have an incentive to share his risk with others, as it may have a detrimental effect to do so. Real world situations where this may happen have been described by several authors.  \cite{hofert2012}  consider nuclear power accidents and estimate the parameters of a Pareto tail to have infinite mean. \cite{eling2019} consider cyber risk and also find parameter values that lead to infinite mean models. Similar observations have already been found for operational risks in \cite{neslehova2006}. 

In recent years several new papers have appeared that study this phenomenon under the assumption of Pareto distributions and more general. \cite{chen-etal2024} have shown that i.i.d. Pareto distributions with infinite mean fulfill the inequality
\begin{equation} \label{def-nodiv}
X_1 \le_{st} \sum_{i=1}^n \theta_i X_i,
\end{equation}
for all $\theta_i \ge 0$, $i=1,\ldots,n,$ with $\sum_{i=1}^n \theta_i =1$. Notice that we can assume without loss of generality that the weights fulfill $\theta_i >0$, as $n$ is arbitrary. We will sometimes use this fact in proofs. This condition means that no rational decision maker will agree to any version of linear risk sharing, see \cite{chen2024risk} for a detailed study of this case. We want to mention that linear peer-to-peer risk sharing rules of this type have recently been studied also in the context of flood risk pooling by \cite{feng2023}. As flood risks can also have very heavy tails this problem may also arise there. 
\cite{chen2024arxiv} have shown that the inequality \eqref{def-nodiv} also holds more general for distributions that they call super-Fr\'echet distributions. Another recent study of linear risk sharing rules can be found in \cite{yang2025}. 

In this paper we want to further study the class of distributions fulfilling the condition \eqref{def-nodiv}. Our main results in section 2 will demonstrate that this condition also holds for all distributions that we will call super Cauchy distributions and we will clarify that this property is strongly related to skewness properties of the distributions relative to a Cauchy distribution.  In section 3 we will consider the case of risks that may have a positive probability of an infinite value with the interpretation of deadly risks. In that case these results about a nondiversification trap are very intuitive and this may help to understand why the situation may be similar in case of very heavy tailed infinite mean risks. Section 4 relates our results to other recent contributions and states some open problems.


{\bf Notation:} Throughout the paper we will assume the existence of a probability space $(\Omega, \mathcal{A}, P)$, on which we can define random variables $X:\Omega \to \mathbb{R}$ with arbitrary distributions. We will denote by 
$
F_X(t) := P(X\le t), \ t \in \mathbb{R},
$
the corresponding cumulative distribution function of $X$, and we will also call $F_X$ just the distribution of $X$. For any distribution function $F$ we define the generalized inverse 
$$
F^{-1}(u) := \inf\{x \in \mathbb{R}: F(x) \ge u\}, \ 0 < u < 1.
$$
We will denote by $X_n \stackrel{d}{\to} X$ convergence in distribution, i.e. pointwise convergence of $F_{X_n}$ to $F_X$ in continuity points of $F_X$. We write $X =_{st} Y$ if $F_X = F_Y$ and we define the usual stochastic order by
$$
X \le_{st}  Y, \ \mbox{ if } F_X(t) \ge F_Y(t) \mbox{ for all } t \in \mathbb{R}.
$$
This usual stochastic order $\le_{st}$ can also be found under the name \textit{first order stochastic dominance} in an economic context and it is generally agreed that any rational decision will prefer a risk $X$ to a risk $Y$, if $X \le_{st}  Y$ holds. It is the strongest of the well known stochastic dominance rules used for decisions under risk and in contrast to weaker notions like the second order stochastic dominance it is defined for any distributions, including infinite mean models. For properties of this and related stochastic orders we refer to the books \cite{muesto2002} and \cite{ShaSha:Springe2007}.

\section{Main results}

Throughout the paper we will consider risks $X$, where we assume that these are random variables on some probability space $(\Omega, \mathcal{A}, P)$ with a known distribution. We allow for arbitrary real values of $X$ even though in many applications, in particular in an insurance context, we may have $X \ge 0$. Even if one is only interested in the case of non-negative risks $X \ge 0$ it is helpful to consider the more general case of arbitrary random variables $X$ as this will give additional insight and more elegant proofs.  In particular it is helpful to relate the problem to stable distributions like the Cauchy distribution, which plays a fundamental role in this context. We will now give the main definition of the class of distributions that we want to consider.
	
\begin{definition}
Let $\mathcal{D}^-$ denote the set of distribution functions $F_X$ with the property that
\begin{equation} \label{def-d}
X_1 \le_{st} \sum_{i=1}^n \theta_i X_i,
\end{equation}
for all $\theta_i \ge 0$, $i=1,\ldots,n,$ with $\sum_{i=1}^n \theta_i =1$, where $X_1,\ldots,X_n$ are i.i.d. random variables with distribution functions $F_X$. We will also write $X \in \mathcal{D}^-$, if  $F_X \in \mathcal{D}^-$.

We denote by $\mathcal{D}^+$ the set of distribution functions $F_X$ where inequality \eqref{def-d} holds in reverse order, i.e.
\begin{equation} \label{def-d+}
\sum_{i=1}^n \theta_i X_i \le_{st} X_1.
\end{equation}
\end{definition}

Notice that $X \in \mathcal{D}^+$ if and only if $-X \in \mathcal{D}^-$. Therefore we will mainly study properties of $\mathcal{D}^-$ in the following. 
The property  $X \in \mathcal{D}^-$ means in terms of risks that diversification of a portfolio of independent such risks has a negative effect by stochastically increasing the risk. This property may be a bit surprising at first sight, as one typically assumes that diversification is good at least for risk averse decision makers. A detailed discussion of this property in the context of risk exchange is given in \cite{chen2024risk}.

The following simple result unifies and generalizes part (ii) - (iv) of Proposition 1 in \cite{chen2024arxiv}.

\begin{theorem} \label{th:main}
Assume that $X,Y \in \mathcal{D}^-$ are independent and that $\phi: \mathbb{R}^2 \to \mathbb{R}$ is an arbitrary increasing convex function. Then  $\phi(X,Y) \in \mathcal{D}^-$.
\end{theorem}

\begin{proof}
This simply follows from the fact that for i.i.d. $X_1, X_2,\ldots$ and $Y_1, Y_2,\ldots$ we get
$$
\phi(X_1,Y_1) \le_{st} \phi\left(\sum_{i=1}^n \theta_i X_i, \sum_{i=1}^n \theta_i Y_i\right) \le \sum_{i=1}^n \theta_i  \phi\left( X_i, Y_i \right).
$$
Here the first inequality follows from \eqref{def-d} for any increasing function as shown in Theorem 1.2.16 of \cite{muesto2002} and it is an immediate consequence of the definition of convexity that the second inequality holds pointwise for all $\omega \in \Omega$. 
\end{proof}

We will now collect a list of several elementary properties of the set $\mathcal{D}^-$. Many of them can be found already in similar forms in the references \cite{chen2024risk,chen-etal2024} and \cite{chen2024arxiv}, but for completeness we add the simple proofs for all of them. 

\begin{theorem} \label{th:elementary}
a) if $X_n \in \mathcal{D}^-, \ n \in \mathbb{N},$  and $X_n \stackrel{d}{\to} X$, then $X \in \mathcal{D}^-$. \\
b) if $X \in \mathcal{D}^-$ is non-degenerate, then $E|X| = \infty$.\\
c)  if $X \in \mathcal{D}^-$, then  $aX +b \in \mathcal{D}^-$ for all $a > 0$ and $b \in \mathbb{R}$.\\
d)  if $X,Y \in \mathcal{D}^-$ and $X,Y$ are independent, then $X + Y \in \mathcal{D}^-$. \\
e)  if $X,Y \in \mathcal{D}^-$ and $X,Y$ are independent, then $\max\{X,Y\}  \in \mathcal{D}^-$. \\
f) if $X \in \mathcal{D}^-$  and $\phi: \mathbb{R} \to \mathbb{R}$ is  increasing convex, then  $\phi(X) \in \mathcal{D}^-$.
\end{theorem}

\begin{proof}
a) This follows immediately from the well known facts, that convergence in distribution is preserved under continuous mappings and thus when building convex combinations, and that $\le_{st}$ is preserved under convergence in distribution, see e.g. Theorem 1.2.14 in \cite{muesto2002}.\\
b) It is well known that $X \le_{st} Y$ and $EX = EY$ implies $X =_{st} Y$, which is an easy consequence of 
$$
EY-EX = \int_{-\infty}^\infty (F_X(t) - F_Y(t)) dt.
$$
Therefore $X \in \mathcal{D}^-$ and  $E|X| < \infty$ implies  $X =_{st} \bar{X}_n := \sum_{i=1}^n X_i/n$ for all $n \in \mathbb{N}$ and due to the law of large number and part a) finally $X =_{st} \lim_{n \to \infty} \bar{X}_n = EX$ a.s. under this assumption.
Part c) - f) immediately follow from Theorem \ref{th:main}.
\end{proof}

We can use Theorem \ref{th:elementary} to show that $X \in \mathcal{D}^-$ implies that this property also holds for scaled limits of sums or maxima of i.i.d. random variables from the corresponding distribution. 
Recall that a random variable $X$ is said to be in the \textit{sum-domain of attraction} of a random variable $Z$, if there exist sequences $a_n > 0$ and $b_n \in \mathbb{R}, \ n \in \mathbb{N}$, such that
\begin{equation} \label{def:doa-plus}
a_n (X_1 + \ldots + X_n)- b_n \stackrel{d}{\to} Z,
\end{equation}
where $X_1,X_2, \ldots$ are i.i.d. copies of $X$. We write $X \in DOA^+(Z)$ in this case. The possible limits are called sum-stable distributions. Similarly, one can define the  \textit{max-domain of attraction} of a random variable $Z$, if there exist sequences $a_n > 0$ and $b_n \in \mathbb{R}, \ n \in \mathbb{N}$, such that
\begin{equation} \label{def:doa-max}
a_n \max\{X_1,\ldots,X_n\} - b_n \stackrel{d}{\to} Z,
\end{equation}
where $X_1,X_2, \ldots$ are i.i.d. copies of $X$. We write $X \in DOA^\vee(Z)$ in the case of the max-domain of attraction of a max-stable random variable $Z$. For details on sum-stable distributions and their domains of attraction we refer to \cite{nolan2020}. The topic of max-stability and their domain of attraction is considered in the extreme value theory literature. We refer to \cite{resnick2008} for a good reference on the details that we will use here. The following result is now easy to show. 

\begin{theorem} \label{th:doa}
a) If $X \in \mathcal{D}^-$ and  $X \in DOA^+(Z)$, then  $Z \in \mathcal{D}^-$. \\  
b) If $X \in \mathcal{D}^-$ and  $X \in DOA^\vee(Z)$, then  $Z \in \mathcal{D}^-$. 
\end{theorem}

\begin{proof}
This follows immediately from the preservation of $\mathcal{D}^-$ under independent sums and maxima, scaling and distributional limits as mentioned  in Theorem \ref{th:elementary}.
\end{proof}

The possible limits in \eqref{def:doa-plus} are necessarily what is called a stable distribution. We use here the following notation for stable distributions, which corresponds to parametrization 1 in \cite{nolan2020}. For $0 < \alpha \le 2$ and $-1\le \beta \le 1$ we say that $Z \sim S(\alpha,\beta)$ has a stable distribution, if the characteristic function fulfills
$$
E(\exp(iuZ)) = \exp(-|u|^\alpha [1-i\beta\tan(\pi \alpha/2)(sign(u))]), \ u \in \mathbb{R}, 
$$
if $\alpha \neq 1$, and 
$$
E(\exp(iuZ)) = \exp(-|u| [1+2i\beta /\pi (sign(u) \ln(|u|))]),\ u \in \mathbb{R}, 
$$
if $\alpha = 1$. Any more general stable distribution with 4 parameters $(\alpha,\beta, \gamma,\delta)$ can be obtained as a linear transformation $\gamma Z + \delta$ for $Z \sim S(\alpha,\beta)$, but for the purpose of this paper it is sufficient to consider stable distributions $Z \sim S(\alpha,\beta)$, as the set 
$\mathcal{D}^-$ is invariant under linear transformations. The parameter $\alpha$ is typically called the characteristic exponent and the parameter $\beta$ is called the skewness parameter. These stable distributions have an infinite mean if $\alpha \le 1$. Any convex combination of i.i.d. stable distributions is again a stable distribution and the obtained parameters can be found in equation (1.7) in \cite{nolan2020}. Using this we can derive the following result for stable distributions.

\begin{theorem} \label{th:stable-d-} 
We have  $Z \sim S(\alpha,\beta)  \in \mathcal{D}^-$ if and only if one of the following conditions holds:\\
a) $\alpha = 1$ and $\beta\ge 0$;\\
b) $\alpha < 1$ and $\beta = 1$.
\end{theorem}

\begin{proof}
For $\alpha > 1$ we have $E|Z| < \infty$ and therefore $Z  \not\in \mathcal{D}^-$. In case $\alpha = 1$ we get from (1.7) in \cite{nolan2020} for any convex combination $\sum w_i X_i$ of i.i.d. stable distributions that $\sum w_i X_i =_{st} X_1 + \delta$, where
$$
\delta = - \frac{2\beta}{\pi} \sum_{i=1}^n w_i \ln(w_i)
$$
and therefore $Z  \in \mathcal{D}^-$ if and only if $\delta \ge 0$, which holds if and only if $\beta \ge 0$. 

If $\alpha < 1$, then we get $\sum w_i X_i =_{st} \gamma X_1$ with 
$$
\gamma = \left( \sum_{i=1}^n w_i^\alpha  \right)^{1/\alpha}  \ge 1,
$$
and thus we have $\gamma > 1$, if $w_i > 0$ for at least two indices $i$. Therefore $\sum w_i X_i \ge_{st} X_1$ in this case, if and only if $P(X_1 \ge 0) = 1$. This holds if and only if $\beta = 1$. 
\end{proof}

Notice that we get a very special case, if we choose $\alpha =1$ and $\beta=0$. Then we get a Cauchy distribution with distribution function
$$
F(x) = \frac{1}{\pi} \arctan(x) + \frac{1}{2}, \ x \in \mathbb{R}.
$$
This has the special property that we get equality in \eqref{def-d}.

Combining Theorem  \ref{th:doa} and Theorem \ref{th:stable-d-} yields a necessary condition for  $X  \in \mathcal{D}^-$, as   $X \in DOA^+(Z)$ and  $Z   \not\in \mathcal{D}^-$ implies 
 $X   \not\in \mathcal{D}^-$. As an example we can consider $X^3$ for $X$ a Cauchy distribution. In this case we get $X \in DOA^+(Z)$  for $Z  \sim S(1/3,0)  \not\in \mathcal{D}^-$ and hence  $X   \not\in \mathcal{D}^-$.
 
Among the max-stable distributions, the Fr\'echet distribution with cdf 
$$
F(x) =  e^{-1/{x^\alpha}}, \ x \ge 0,
$$
has an infinite mean, if $\alpha \le 1$. A random variable $X$ in the sum-domain of attraction of a stable distribution $Z$ with parameter $\alpha \le 1$ fulfills  $X \in DOA^\vee(Z)$ for a Fr\'echet distribution $Z$ with the same parameter $\alpha$. This follows from the fact that both domains of attraction are characterized by regular variation of order $\alpha$ for the survival function, see Theorem 3.14 in \cite{nolan2020} and Proposition 1.11 in \cite{resnick2008} for details. Therefore we can derive the following result. 

\begin{theorem} \label{th:max-stable-d-} 
We have $X \in DOA^\vee(Z)$ with  $Z  \in \mathcal{D}^-$ if and only if $Z$ is a Fr\'echet distribution with $\alpha \le 1$.
\end{theorem}

\begin{proof}
According to Theorem \ref{th:stable-d-} we have for any $\alpha \le 1$ a stable distribution $X \in \mathcal{D}^-$. If $X \in DOA^\vee(Z)$, then it follows from Theorem \ref{th:doa} that $Z \in \mathcal{D}^-$. But this holds for $Z$ a Fr\'echet distribution with $\alpha \le 1$. As these Fr\'echet distributions are the only max-stable distributions with infinite mean, the result follows. 
\end{proof}

\cite{chen2024arxiv} gave an alternative direct proof of the result that infinite mean Fr\'echet distributions are in  $\mathcal{D}^-$. We will demonstrate now that this result as well as the  related results for Pareto distributions given in \cite{chen-etal2024} are all special cases of the simple fact that all these distributions are convex transformations of Cauchy distributions. 
Notice that for continuous distributions $Y = \phi(X)$ can only hold for an increasing $\phi$, if we choose $\phi(x) = F_Y^{-1}(F_X(x))$. This relative inverse distribution function $\phi$ is strongly related to the statistical concept of a quantile-quantile plot (or shortly Q-Q plot). The condition of the relative inverse being convex is a well known order relation between distributions describing an increase in skewness. This concept is discussed in detail in \cite{zwet1964} and \cite{oja1981}. We will use the following notation.

\begin{definition} \label{skew-order}
A cdf $G$ is said to be more skewed than $F$ (written $F \le_{skew} G$), if the function $\phi(x) = G^{-1}(F(x))$ is convex. 
\end{definition}

If densities $f$ and $g$ exist, then taking the derivative of $\phi$ yields the condition that $F \le_{skew} G$
holds if and only if
\begin{equation} \label{skew-density}
u \mapsto \frac{f(F^{-1}(u))}{g(G^{-1}(u))}  \quad \mbox{ is non-decreasing, } u \in (0,1). 
\end{equation}

In the case of a Cauchy distribution, where we get equality in distribution in \eqref{def-d}, we can even choose arbitrary convex transformations to remain in $\mathcal{D}^-$.

\begin{theorem} \label{th:main-cauchy}
Assume that $X$ has a Cauchy distribution and that $\phi: \mathbb{R} \to \mathbb{R}$ is an arbitrary convex function. Then  $\phi(X) \in \mathcal{D}^-$.
\end{theorem}

\begin{proof}
This simply follows from the fact that convexity of $\phi$ implies for any convex combination
$$
\phi(X_1) =_{st} \phi\left(\sum_{i=1}^n \theta_i X_i\right) \le \sum_{i=1}^n \theta_i  \phi\left( X_i\right) 
$$
P.-a.s..
\end{proof}

For a distribution $F$ with a heavy right tail,  a convex transformation leads to a distribution, which has even more heavy right tails. Therefore several authors introduced classes of \textit{super heavy-tailed distributions} by requiring a convex transformation of a specific heavy-tailed distribution. We will consider now the relationship between some of these concepts and their relation to the class $\mathcal{D}^-$.

\begin{definition}
a) For any c.d.f. $F$ we can define the class of distributions
$$
\mathcal{S}(F) := \{G: F \le_{skew} G\}.
$$
b) If 
$$
F(x) =  1 -  \frac{1}{x}, \ x \ge 1,
$$
is the c.d.f. of a Pareto(1) distribution, then $\mathcal{S}_P := \mathcal{S}(F)$ is called the class of \textit{super Pareto distributions}. 

If 
$$
F(x) =  e^{-1/x}, \ x \ge 0,
$$
is the c.d.f. of a standard Fr\'echet distribution, then $\mathcal{S}_F := \mathcal{S}(F)$ is called the class of \textit{super Fr\'echet distributions}. 

If
$$
F(x) = \frac{1}{\pi} \arctan(x) + \frac{1}{2}, \ x \in \mathbb{R},
$$
is the c.d.f. of a Cauchy distribution, then $\mathcal{S}_C := \mathcal{S}(F)$ is called the class of \textit{super Cauchy distributions}. 
\end{definition}

The class of super Pareto distributions has been introduced in \cite{chen-etal2024} and the class of super Fr\'echet distributions was considered in \cite{chen2024arxiv}. It is natural to also consider the class of super Cauchy distributions and we will show now the relationsship between these classes and that they are all contained in $\mathcal{D}^-$. 

\begin{theorem} \label{main-relations}
The following relations hold: $\mathcal{S}_P \subset \mathcal{S}_F \subset \mathcal{S}_C \subset \mathcal{D}^-$.
\end{theorem}

\begin{proof} The relationship $\mathcal{S}_P \subset \mathcal{S}_F$ follows from Example 11 in \cite{chen2024arxiv}, where it is shown that a Pareto(1) distribution is super Fr\'echet. Therefore we mainly have to show $\mathcal{S}_F \subset \mathcal{S}_C$. 
Consider a Fr\'echet distribution with $G(x) = e^{-1/x}$ and a Cauchy distribution $F(x) =  \frac{1}{\pi} \arctan(x) + \frac{1}{2}$. We have to show that $G^{-1}(F(x))$ is convex, or equivalently that $h(x) = F^{-1}(G(x))$ is concave. We get 
$$
h(x) = \tan(\pi e^{-1/x} - \pi/2), \ x > 0. 
$$
Taking the second derivative yields
$$
h''(x) = - \frac{\pi e^{-2/x} (e^{1/x} (2 x -1) + 2 \pi \cot(\pi /e^{1/x})) }{ \sin^2(\pi e^{-1/x}) x^4}.
$$
For $x \le \pi/4$ the slope of $\tan(x)$ is smaller than $2$ and thus we have $\tan(x) \le 2x$ and hence $\cot(x) \ge 1/(2x)$. This implies $h''(x) \le 0$ for $x \le 1/2 \le \pi/4$.  For $x \ge 1/2$ it obviously holds  that $h''(x) \le 0$. Hence $h$ is concave and thus $F$ is less skewed than $G$. The relation $\mathcal{S}_C \subset \mathcal{D}^-$ is an immediate consequence of Theorem \ref{th:main}.
\end{proof}

This result shows that there are additional distributions in  $\mathcal{D}^-$ compared to the classes considered in  \cite{chen-etal2024} and \cite{chen2024arxiv} .

We will now show that even the restriction of $\mathcal{S}_C$ to non-negative distributions is strictly larger than $\mathcal{S}_F$.

\begin{theorem}
If $X$ has a Cauchy distribution, then the distribution function $F(x) = 2\arctan(x)/\pi, \ x \ge 0$,  of $|X|$ fulfills $F \in \mathcal{S}_C$, but $F \not\in \mathcal{S}_F$.
\end{theorem}

\begin{proof}
To show that this distribution with $F(x) = 2\arctan(x)/\pi$ is not a super Fr\'echet distribution, we have to consider the relative inverse with respect to $G(x) =e^{-1/x}$. We get
$$
\phi(x) = F^{-1}(G(x)) = \tan(\pi e^{-1/x}/2).
$$
A straightforward calculation similar to the one in Theorem \ref{main-relations} shows that this function is concave for large $x$ and thus $F \not\in \mathcal{S}_F$.
\end{proof}

\begin{theorem}
If $X \ge 0$ is non-degenerate, then $X \not\in \mathcal{D}^+$.
\end{theorem}

\begin{proof}
Assume that $X \ge 0$ is non-degenerate and $X \in \mathcal{D}^+$ so that in particular
\begin{equation} \label{eq:cauchy}
X \ge_{st} \frac{1}{n} \sum_{i=1}^n X_i =: Y_n
\end{equation}
for $X_1,X_2, \ldots$ i.i.d. with the same distribution as $X$ and for all $n \in \mathbb{N}$. As the Cauchy distribution is the only non-degenerate distribution with stochastic equality in \eqref{eq:cauchy} for all $n$, we must have $F_X(t_0) < F_{Y_n}(t_0)$ for some $t_0 > 0$ and for some $n \in \mathbb{N}$, if $X \ge 0$ fulfills \eqref{eq:cauchy}.  For $t > t_0$ we get
$$
E(t-X)_+ = \int_0^t F_X(z) dz < \int_0^t F_{Y_n}(z) dz = E(t-Y_n)_+. 
$$
However, if $X$ has a finite mean, then we have $E[X_1|Y_n] = Y_n$ and thus $X \ge_{cx} Y_n$. Therefore we get for the convex function $f(x) = (t-x)_+$ that
$$
Ef(X) = E(t-X)_+ \ge E(t-Y_n)_+,
$$
a contradiction. If $X$ does not have a finite mean, this still applies, as the proof is only based  on values of $X$ below $t$, and therefore we can apply the argument to the bounded random variable $\tilde{X} := \min\{X,t\}$. 
\end{proof}

We want to mention, that there are of course also random variables with $X \ge 0$, $EX=\infty$ and $X \not\in \mathcal{D}^-$. The following example demonstrates that $X \in \mathcal{D}^-$ leads to conditions also for the values of the c.d.f. for small $x$ and thus also demonstrates that there can not be any sufficient conditions for $X \in \mathcal{D}^-$ based only on the tail behavior of the distribution of $X$.

\begin{example}
Assume that $X_1, X_2$ are i.i.d. with $X_1 \ge 0$, $EX_1 = \infty$ and $P(0 \le X_1 \le 2) = P(2 < X_1 \le 4) = 0.4$. Then
$$
P((X_1+X_2)/2 \le 3) \ge P(0 \le X_1 \le 4,0 \le X_2 \le 2) +  P(0 \le X_1 \le 2,2 < X_2 \le 4) = 0.48.
$$
Hence $F \not\in \mathcal{D}^-$, if $F(0)=0, F(2) = 0.4, F(3) < 0.48$ and $F(4) = 0.8$. 

Notice that we can derive similar and sharper bounds for Super-Pareto and related distributions for $F(3)$, if we assume $F(0)=0, F(2) = 0.4$ and $F(4) = 0.8$. If we denote by $G$ the distribution function of a Pareto(1) distribution, then $F$ is Super-Pareto, if and only if $G^{-1}(F(x))$ is concave. This implies in case $F(0)=0, F(2) = 0.4$ and $F(4) = 0.8$ that
$$
G^{-1}(F(3)) \ge \frac{1}{2} ( G^{-1}(F(2))  + G^{-1}(F(4))  )  = \frac{10}{3}
$$
and thus $F(3) \ge 0.7$. A similar calculation yields that in the case of a super-Fr\'echet distribution we get the condition $F(3) \ge 0.698$ and in the case of a super-Cauchy distribution we get the necessary condition $F(3) \ge 0.654$.
\end{example}

\section{Deadly risks with $P(X= \infty) > 0$}

Let us now consider the case that the random variables have values in $\mathbb{R} \cup \{\infty\}$. In particular we consider the case of $P(X_1 = \infty) = p = 1-P(X_1=0)$ for some $p > 0$. Then it obviously holds that for a convex combination
$$
Y = \sum_{i=1}^n \theta_i X_i \ \mbox{ we get } P(Y = 0)  = \prod_{i=1}^n (1-p) 1_{[\theta_i > 0]} = 1-P(Y = \infty),
$$
and hence this distribution is in $\mathcal{D}^-$. Notice that we can write $X_i = \phi(Z_i)$ for any continuous real valued random variable $Z$ by choosing
$$
\phi(x) = 
\left\{
\begin{array}{cl}
\infty &, \mbox{ if } x \ge F_X^{-1}(p),\\
0 &, \mbox{ else.}
\end{array}
\right.
$$
This function $\phi$ is convex, so that we can consider this distribution as the extreme case of the most skewed infinity mean distribution that exists. We can give a nice intuitive interpretation of this result by assuming that $X = \infty$ means a deadly risk. Assume that 
a pool of people stranded on a deserted island. Every individual collects some food.
You have to eat a certain amount of food to survive, but all available items have independent of the others a probability $p$ of being poisoned and leading to death even if you only eat a small amount of it. So if you must eat something to survive, it is intuitively clear  that it is optimal to just eat from one item leading to a probability of death of $p$, whereas you can only increase this probability if you eat from several potentially deadly items. In this case we have the property that $X_1$ and $c X_1$ have the same distribution for every $c > 0$ and this makes it immediately obvious that such a result holds. So in this case of deadly risks it is intuitively clear that risk sharing is not a good idea. Knowing this it may appear a bit less surprising that similar things can happen in the case of infinite mean models with extremely heavy tails.

\section{Relation to other recent contributions and open problems.}

Several other authors recently considered this question of finding large classes of distributions in $\mathcal{D}^-$ independently of our approach. 
\cite{chen2024arxiv} define the class $\mathcal{H}$ of distributions with the property that $x \mapsto h_F(x) := - \log(F(1/x))$ is subadditive. 
\cite{arab2024} consider the class of distributions with the property that $x \mapsto g_F(x) := 1-F(1/x)$ is subadditive. Let us call this class of distributions by $\mathcal{G}$. It is easy to see that $\mathcal{H} \subset\mathcal{G}$. Both papers contain examples of distributions $F$ in their respective class with $F \not\in \mathcal{S}_C$. A simple example with $F \in \mathcal{H}$, but $F \not\in\mathcal{S}_C$ can be constructed by using a subadditive function $h_F$ with a kink with increasing slope. As a concrete example we can choose
$$
h_F(x)  = 
\left\{
\begin{array}{rl}
2x, & x < 1,\\
2, & 1 \le x < 2,\\
x, & x \ge 2. 
\end{array}
\right.
$$
As $x \mapsto h_F(x)/x$ is obviously decreasing, this is a subadditive function. The corresponding distribution function $F$ fulfills $F(x) = \exp(-h_F(1/x))$ and therefore also has a kink with increasing slope. But this implies that $F \not\in\mathcal{S}_C$.

  We give now an example of a non-negative distribution with $F \in \mathcal{S}_C$  with $F \not\in \mathcal{G}$. We consider the following convex function $\phi: \mathbb{R} \to (0,\infty)$: 
$$
\phi(x) :=
\left\{
\begin{array}{rl}
\frac{1}{10-x}, & x < 5,\\
\frac{x}{25}, & x \ge 5.
\end{array}
\right.
$$
If $X$ has a Cauchy distribution, then we get for the distribution $F$ of $\phi(X)$ that 
$$
g_F(x)  = \frac{1}{2} - \frac{1}{\pi} \cdot
\left\{
\begin{array}{rl}
\arctan(25/x), & x < 5,\\
\arctan(10-x), & x \ge 5.
\end{array}
\right.
$$
Thus we have $g_F(10) = 0.5 > 2\cdot g_F(5) \approx 0.12$ and thus $g_F$ is not subadditive and hence $F \not\in \mathcal{G}$. 

\cite{chen2024arxiv}  claim in their Example 3 that there is an example of a discrete distribution $F \in \mathcal{H}$. Unfortunately, this example is wrong, as the corresponding function $h_F$ is not subadditive. Indeed it is easy to see that a subadditive function $h_F$ with $\lim_{x \to 0} h_F(x) = 0$ must be continuous and thus any distribution in $\mathcal{H}$ also must be continuous. This also holds for distributions in $\mathcal{G}$. The example of a discrete distribution in $\mathcal{G}$ mentioned in \cite{chen2024arxiv} has positive probability $P(X=\infty) > 0$ and this is similar to the example that we consider in the previous section.  It seems to remain an open problem, whether there is a discrete distribution in $\mathcal{D}^-$ that assumes only finite values.

It also seems to be an open problem whether all stable distributions in $\mathcal{D}^-$ are super Cauchy distributions. 

For stable distributions with parameter $\alpha < 1$ and $\beta =1$ some numerical experiments seem to support this conjecture but we have no idea how to prove it, as an explicit formula for the cdf is only known for the case $\alpha =1/2$, which is known as Levy distribution.  So the question arises whether one can prove that these positive infinite-mean stable distributions are convex transformations of a Cauchy distribution for any $\alpha < 1$ without having a closed form for the cdf? 

A similar problem arises for  stable distributions with parameter $\alpha = 1$ and $\beta > 0$. We know from Theorem \ref{th:stable-d-}  that these distributions are also in $\mathcal{D}^-$, but even though the parameter $\beta$ is commonly denoted as skewness parameter, we could not find a formal proof, that an increase in $\beta$ yields an increase with respect to the stochastic order $\le_{skew}$ for stable distributions. We also conjecture that this is true in general. 

\textbf{Competing interests}: the author declares no competing interests. 

\textbf{Acknowledgement}: I thank the editors and two anonymous referees for several helpful remarks that considerably improved the paper. 
	
	\bibliographystyle{apalike}
	\bibliography{stable}
	
	\appendix
\end{document}